\documentclass[apj]{emulateapj}

\newcommand{\ratio}{_{3-2/1-0}}
\newcommand{\cott}{_{{\rm CO}(3-2)}}
\newcommand{\cooz}{_{{\rm CO}(1-0)}}
\newcommand{\kkms}{K\,km\,s$^{-1}$}
\newcommand{\kms}{km\,s$^{-1}$}

\usepackage[usenames]{color}

\shorttitle{Enhancement of CO($3-2$)/CO($1-0$) Ratios and SFE in GHRs}

\shortauthors{R. E. Miura et al.}

\begin{document}


\title{Enhancement of CO($3-2$)/CO($1-0$) Ratios and Star Formation Efficiencies in Supergiant H{\sc ii} Regions}


\author{Rie E. Miura\altaffilmark{1,2}, Kotaro Kohno\altaffilmark{3,4}, Tomoka Tosaki\altaffilmark{5}, Daniel Espada\altaffilmark{1,6,9}, Akihiko Hirota\altaffilmark{7}, Shinya Komugi\altaffilmark{1}, Sachiko K. Okumura\altaffilmark{8}, Nario Kuno\altaffilmark{7,9}, Kazuyuki Muraoka\altaffilmark{10}, Sachiko Onodera\altaffilmark{10}, Kouichiro Nakanishi\altaffilmark{1,6,9}, Tsuyoshi Sawada\altaffilmark{1,6}, Hiroyuki Kaneko\altaffilmark{11}, Tetsuhiro Minamidani\altaffilmark{7}, Kosuke Fujii\altaffilmark{1,2}, Ryohei Kawabe\altaffilmark{1,6}}
\altaffiltext{1}{National Astronomical Observatory of Japan, 2-21-1 Osawa, Mitaka, Tokyo, 181-8588, Japan}
\email{rie.miura@nao.ac.jp}
\altaffiltext{2}{Department of Astronomy, The University of Tokyo, Hongo, Bunkyo-ku, Tokyo, 133-0033, Japan}
\altaffiltext{3}{Institute of Astronomy, School of Science, The University of Tokyo, Osawa, Mitaka, Tokyo 181-0015, Japan}
\altaffiltext{4}{Research Center for Early Universe, School of Science, The University of Tokyo, Hongo, Bunkyo, Tokyo, 113-0033, Japan}
\altaffiltext{5}{Joetsu University of Education, Yamayashiki-machi, Joetsu, Niigata, 943-8512, Japan} 
\altaffiltext{6}{Joint ALMA Observatory, Alonso de Cordova 3107, Vitacura 763-0355, Santiago de Chile}
\altaffiltext{7}{Nobeyama Radio Observatory, Minamimaki, Minamisaku, Nagano, 384-1805, Japan}
\altaffiltext{8}{Department of Mathematical and Physical Sciences, Faculty of Science, Japan Womanfs University, Mejirodai 2-8-1, Bunkyo, Tokyo 112-8681, Japan}
\altaffiltext{9}{Department of Astronomical Science, The Graduate University for Advanced Studies (Sokendai), 2-21-1 Osawa, Mitaka, Tokyo 181-0015, Japan}
\altaffiltext{10}{Osaka Prefecture University, Gakuen 1-1, Sakai, Osaka 599-8531, Japan}
\altaffiltext{11}{Department of Physics, Meisei University, Hino, Tokyo 191-8506, Japan}
\altaffiltext{12}{Graduate School of Pure and Applied Sciences, Institute of Physics, University of Tsukuba, 1-1-1 Tennodai, Tsukuba, Ibaraki 305-8571, Japan}


\begin{abstract}
We present evidence that super giant H{\sc ii} regions (GHRs) and other disk regions of the nearby spiral galaxy, M33, occupy distinct locations in the correlation between molecular gas, $\Sigma_{\rm H_2}$, and the star formation rate surface density, $\Sigma_{\rm SFR}$.
This result is based on wide field and high sensitivity CO($3-2$) observations at 100\,pc resolution.
Star formation efficiencies (SFE), defined as $\Sigma_{\rm SFR}/\Sigma_{\rm H_2}$, in GHRs are found to be $\sim1$ dex higher than in other disk regions.  
The CO($3-2$)/CO($1-0$) integrated intensity ratio, $R\ratio$, is also higher than the average over the disk.
Such high SFE and R$\ratio$ can reach the values found in starburst galaxies, which suggests that GHRs may be the elements building up a larger scale starburst region.
Three possible contributions to high SFEs in GHR are investigated: (1) the $I_{\rm CO}-N({\rm H}_2)$ conversion factor, (2) the dense gas fraction traced by $R\ratio$, and (3) the initial mass function (IMF).
We conclude that these starburst-like properties in GHRs can be interpreted by a combination of both a top-heavy IMF and a high dense gas fraction, but not by changes in the $I_{\rm CO}-N({\rm H}_2)$ conversion factor.
\end{abstract}


\section{Introduction}
Large starburst systems are characterized by high star formation rates (SFR) \citep{1998ApJ...498..541K}.
Recently, \citet[][hereafter D10]{2010ApJ...714L.118D} have suggested the existence of two different star formation (SF) regimes based on the so-called SF law \citep[][]{1959ApJ...129..243S,1989ApJ...344..685K}, the relation between the molecular gas ($\Sigma_{\rm H_2}$) and SFR surface densities ($\Sigma_{\rm SFR}$): {\it i}) a rapid mode of SF for starbursts, and {\it ii}) a long-lasting mode for disks.
While the former mode is mostly found in active star forming galaxies, having about one dex higher SF efficiencies (SFE, defined as $\Sigma_{\rm SFR}/\Sigma_{\rm H_2}$), the latter is composed by relatively normal galaxies.
It has been claimed that the difference of these two regimes may be due to the different dense gas fractions, the effect of a top-heavy initial mass function (IMF) in starbursts (D10), or to the $I_{\rm CO}-N({\rm H}_2)$ conversion factor \citep[hereafter $X_{\rm CO}$ factor; e.g.,][]{2011ApJ...740L..15M}.

Although not following a bimodal distribution, a large variation in SFE is known on resolved scales from hundreds of parsec to parsec scale in our Galaxy and other galaxies \citep[e.g.][]{2010ApJ...722.1699S,2010ApJ...722L.127O, 2010ApJ...723.1019H, 2012ApJ...745..190L,2013AJ....146...19L,2013ApJ...778..133L}.
\citet{2013AJ....146...19L} suggested that the large scatter in SFE (0.3 dex) at kpc resolution in nearby galaxies.
These can be attributed to different $X_{\rm CO}$ factors.
These studies are still beyond the unit of SF, i.e., Giant Molecular Cloud (GMC), and their measurements average over kiloparsec scales including areas with very different physical conditions.
At resolved GMC scales (several tens of parsec) the scatter on the SF law becomes even larger, which might be due to the spatial offsets between star forming regions and molecular clouds \citep{2010ApJ...722.1699S,2010ApJ...722L.127O,2010ApJ...721.1206C}.

At even higher resolution such as molecular clump scale (parsec scale), the observed SFE for Galactic dense clumps exceeds the observed extragalactic predictions by factors of 17--50 \citep{2010ApJ...723.1019H}.
Such large variations in SFE can be attributed to the different volume density of individual clumps \citep{2010ApJ...723.1019H, 2012ApJ...745..190L}. 
However these Galactic studies corresponds to low mass star forming regions and do not elucidate the physical conditions of massive star forming regions, which is the main source in terms of energy output in starburst galaxies.

The main goal of this Paper is to investigate the underlying relation between molecular gas and SF at GMC scales, especially focusing on the massive SF.
This allow us to infer the properties of the building elements of starbursts.
Super giant H{\sc ii} regions (GHRs), characterized by a H$\alpha$ luminosity of more than $10^{39}$\,erg\,s$^{-1}$ and a size of more than 100\,pc \citep{1989ApJ...337..761K}, are arguably among the most important candidates to understand the SF process occurring at distant starburst systems because there is a close resemblance of properties between them.
In fact, GHRs are often called ``mini-starbursts'' because they contain tens to hundreds of young clusters rich in O, B, Wolf-Rayet stars and are predicted to have quasi-instantaneous events of SF in a few Myr \citep[e.g.,][]{2006AJ....131..849P}.
The majority of massive stars born in GHRs are expected to be formed inside 100\,pc-scale giant molecular clouds \citep[GMC;][]{1996AJ....111.1252M}.

Nearby extragalactic GHRs not only permit detailed studies of individual stars and their parent molecular gas but also of the SF law under these peculiar environmental conditions.
M33 is the nearest spiral galaxy with the most luminous GHRs in the Local Group galaxies (LGGs), which makes it one of the best laboratories to study this problem.
Its favourable inclination is also ideal to resolve the typical size of GMCs with less contamination of projected emission along the line of sight, unlike in our Galaxy.
Another advantage over our Galaxy is that all GMCs are at the same distance.

We obtained wide field and high resolution single-dish mapping of the CO($3-2$) emission toward M33 \citep[][hereafter Paper\,{\sc i}]{2012ApJ...761...37M}, as part of the NRO MAGiC project \citep{2011PASJ...63.1171T,2011PASJ...63.1139K,2012PASJ...64..133O}.
In Paper\,{\sc i}, we identified 65 GMCs and classified them into four categories according to their spatial correlation with young ($<10$--30\,Myr-old) stellar groups (YSGs) and H{\sc ii} regions.
This classification was interpreted as an evolutionary sequence of GMCs.

In this Paper we aim (1) to check whether the SF law holds around GHRs, (2) to compare the SF law with that in large scale starburst environments, and (3) to infer the origin of any peculiarity of the molecular gas properties and SF in GHRs.
Note that although previous molecular SF law studies in M33 have been carried out using different CO transition lines at 80--200\,pc resolution \citep[e.g.,][]{2004ApJ...602..723H, 2010ApJ...722L.127O, 2012PASJ...64..133O}, a detailed SF law taking into account these peculiar regions has not been performed so far.

\section{Data}
\label{observation}
\subsection{Molecular gas surface densities}
\begin{figure}
\plotone{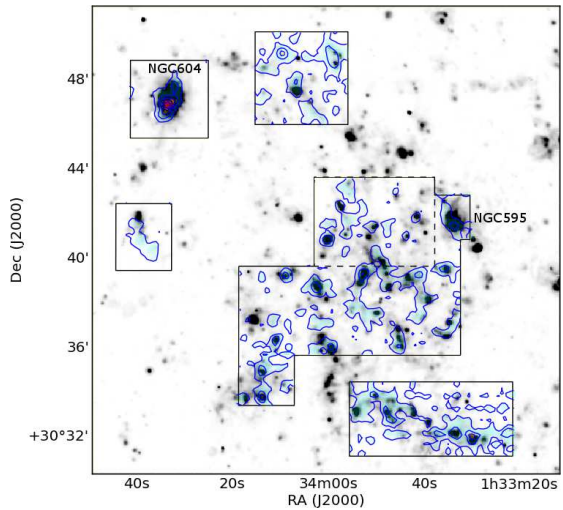}
\caption{The CO($3-2$) integrated intensity map of M33 (filled contour), overlaid on the grey scale SFR image. 
Contour levels are 1, 3, 5, 7, and 9\,\kkms. The black boxes represent the observed regions in CO($3-2$) emission presented in Paper {\sc i}. The dashed box is the newly observed region. \label{fig1}}
\end{figure}

We use ASTE 10-m CO($3-2$) and NRO 45-m CO($1-0$) data cubes.
See Paper\,{\sc i} for a description of the observations and data reduction.
The observed regions are shown in Figure\,\ref{fig1}.
However, note that in this Paper we have added CO($3-2$) data for the northern inner kiloparsec region ($325\arcsec\times240\arcsec$; see Figure\,\ref{fig1}). 
The new observations were performed between November and December 2011, using the ASTE 10-m dish \citep[][]{2004SPIE.5489..763E,2008SPIE.7012E...6E}.
The On-The-Fly mapping technique was employed to obtain the CO($3-2$) data.
The main beam efficiency is measured  to be $0.5\,\pm\,0.1$.
The typical system temperatures in a single side band were 300 to 400\,K. 
The spatial resolution of the final map is 25$\arcsec$ ($\sim100\,$pc) and the grid spacing 8$\arcsec$.
The data calibration and reduction were performed in a similar way to the previously obtained data for other fields of views (Paper\,{\sc i}). Further information on these new ASTE observations will be reported in detail in a forthcoming paper.

The rms per 2.5\,\kms\ velocity resolution ($\sigma_{\rm ch}$) spans 16--32\,mK, depending on the regions.
The CO($3-2$) and CO($1-0$) integrated intensity maps are created over the same velocity range where emission was above 2$\sigma_{\rm ch}$.
The rms of the integrated intensity maps ($\sigma_{\rm mom}$) for each region is 0.16--0.88 \kkms\ and 0.72--1.30 \kkms, for the CO($3-2$) and CO($1-0$) maps, respectively. Thus the sensitivity of CO($3-2$) data is better by a factor of 1--5 (depending on the regions) than that of the CO($1-0$) data.

\subsection{Star Formation Rate}
We use the same calibrated H$\alpha$ image and the {\it Spitzer} 24\,$\micron$ data as in Paper\,{\sc i}.
In this Paper, we have applied a local background subtraction \citep[e.g.][]{2011ApJ...735...63L} for these data sets
using the {\it HIIphot} package \citep{2000AJ....120.3070T}.
The extinction-corrected H$\alpha$ map was created using a linear combination of the two local background subtracted luminosities, and then the SFR per unit area was calculated using the equation\,(7) in \citet{2007ApJ...666..870C}.
The resultant SFR image at 5$\farcs$7 resolution is shown in Figure\,\ref{fig1}.
The resulting rms noise of the SFR maps is $6.5\times10^{-5}$ M$_{\sun}$\,yr$^{-1}$\,kpc$^{-2}$, corresponding to an uncertainty of 4\,\% on average. 
Finally the SFR image was convolved and regridded to a common angular resolution, 25$\arcsec$, so that each data point corresponds to a resolution element when we do a pixel-to-pixel analysis of the SF law.

\section{Star Formation Law in Giant H{\sc ii} Regions of M33}
\label{sflaw}
First, we calculate the best linear fit in the form $\log \Sigma_{\rm SFR} [M_{\sun}\,{\rm yr^{-1}\,kpc}^{-2}]= \alpha \log I_{{\rm CO}(3-2)} [\rm K\,km\,s^{-1}]+\beta$ to all the data with $I_{\rm CO(3-2)} > 2\sigma_{\rm mom}$, and we obtain $\alpha=1.04\pm0.14$ and $\beta=-2.21\pm0.03$. 
We use the ordinary least-squares (OLS) bisector fit following previous studies of the SF law \citep[e.g.,][]{2008AJ....136.2846B} to be able to compare.
The slope in the $\Sigma_{\rm SFR}$-$I_{\rm CO(3-2)}$ plot of M33 is close to unity as found in other studies \citep[][]{2009ApJ...695.1537I,2012MNRAS.424.3050W,2007PASJ...59...55K}.
Our result confirms that the SF law with CO($3-2$) is kept from GMC scales (100\,pc) to large scales, over a variety of environments and physical conditions.

\begin{figure}
\plotone{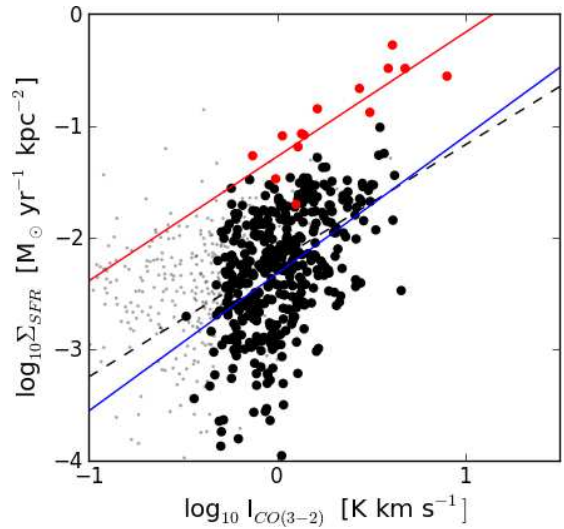}
\caption{Comparison between the CO($3-2$) intensities and SFR surface densities for the GHRs (red) and for the non-GHR (black), in a logarithmic scale. Data points where the intensity is lower than $2\sigma$ are shown in gray. The OLS fits to the data ($>2\sigma$) for all regions, GHRs and non-GHRs are shown as a dashed line, red, and blue solid lines, respectively.\label{fig2}}
\end{figure}

Next we investigate the SF law for regions close to GHRs.
We present a $\Sigma_{\rm SFR}$-$I_{\rm CO(3-2)}$ plot in Figure\,\ref{fig2}.
We compare regions with GHRs and without GHRs.
We restrict to radii within 200\,pc from the centre of the GHR to probe molecular gas potentially affected by the GHRs \citep{1997ApJ...483..210W}.
Our CO($3-2$) observed regions include two of the most luminous GHRs in LGGs, NGC~595 and NGC~604.
The bisector fit for regions close to GHRs is $\alpha=1.11\,\pm\,0.21$ and $\beta=-1.28\,\pm\,0.09$ and for the non-GHRs $\alpha=1.23\,\pm\,0.12$ and $\beta=-2.33\,\pm\,0.03$.
While the slopes are comparable within the uncertainties, we find that there is a remarkable offset in the intercept of $\sim$ 1 dex.

\begin{figure}
\plotone{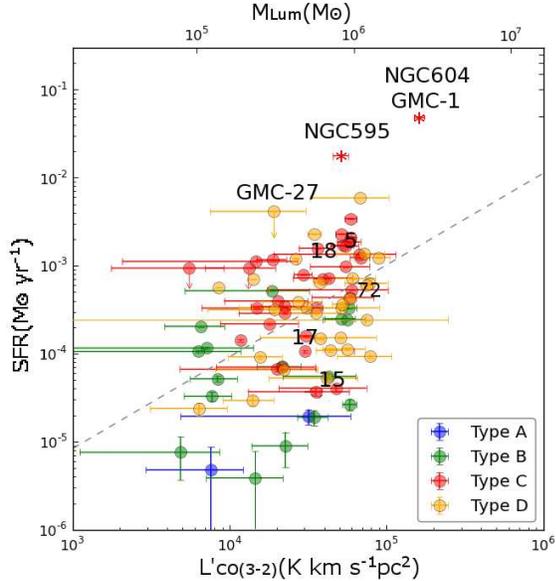}
\caption{
Comparison between the CO($3-2$) luminosities and SFR surface densities for the two GMCs associated with GHRs, NGC\,604 and NGC\,595, as well as M33 GMCs in the four different evolutionary stages from Paper\,{\sc i} are indicated in color code.
The dashed line represents the best fit to all these data points.
The GMCs used in the LVG-analysis in Figure\,\ref{fig6a} are labeled: GMC-1 (associated with the GHR NGC\,604), GMC-5, 15, 17, 18, 27 and 72. \label{fig2a}}
\end{figure}

Because this is a pixel-to-pixel plot with $\sim100$\,pc scale,
this offset might be caused by the drift of young clusters from their parent GMCs as suggested by \citet{2010ApJ...722.1699S} and \citet{2010ApJ...722L.127O}.
In order to check if this possibility contributes to the difference in SFR between GHR and non-GHR points, we plot in Figure\,\ref{fig2a} the SFR and CO luminosities averaged over each GMC from our CO(3--2) GMC catalog (Paper\,{\sc i}; Miura et al. in preparation).
Note that most of the young star forming regions associated with each GMC are still within the GMC boundary (Paper\,{\sc i}).
The star symbols are the data points for GHRs, while the other filled circles represent each of the four evolutionary stages of GMCs in M33:
Type\,A GMCs show no sign of massive SF, Type\,B are associated only
with relatively small H{\sc ii} regions, Type\,C with both H{\sc ii} regions and relatively young ($<10$\,Myr) YSGs,
and Type\,D with both H{\sc ii} regions and relatively old ($>10$\,Myr old) YSGs (Paper\,{\sc i}).

We find a large scatter in the fit for the M33 GMCs.
Although the SF law is maintained on GMC scales, part of the scatter
may arise from differences in the evolutionary stages of the GMCs. 
The data points for Type\,A and B are located preferentially below the fit, while those for Type C and Type D, distribute around the fitted line.
The data points for NGC\,604 and NGC\,595 are still 1dex higher than the ones for non-GHRs, even in the SF law for individual GMCs.
This suggests that high SFEs are not caused by an offset between the molecular gas and the SF regions.

We note that GMC-27 also seems to have a relevant offset with respect to the fit.
GMC-27 is located at the vicinity of the GHR NGC 604, but it was distinguishable
from the nearby GMC associated with NGC 604 by a large velocity difference of
20 km\,s$^{-1}$ \citep{2010ApJ...724.1120M}. It is necessary to avoid the possible contamination in the star
formation tracers (H$\alpha$ emission and 24\,$\micron$) along the line of sight, but it is not possible due to lack of velocity information. Thus the relatively high SFR at GMC-27 is
likely an overestimation due to contamination by the neighbouring GHR.

\section{Comparison with Other Galaxies}
A bimodal behavior of rapid vs slow SF modes was found in external galaxies \citep[D10;][]{2005PASJ...57..733K}: {\it normal} star forming galaxy such as spiral and BzK galaxies \citep{2004ApJ...617..746D}, and starburst galaxies such as luminous infrared and submillimeter galaxies, respectively.
In this section, we compare the SF law for the GHRs in M33 with that in the external galaxies.

\begin{figure}
\plotone{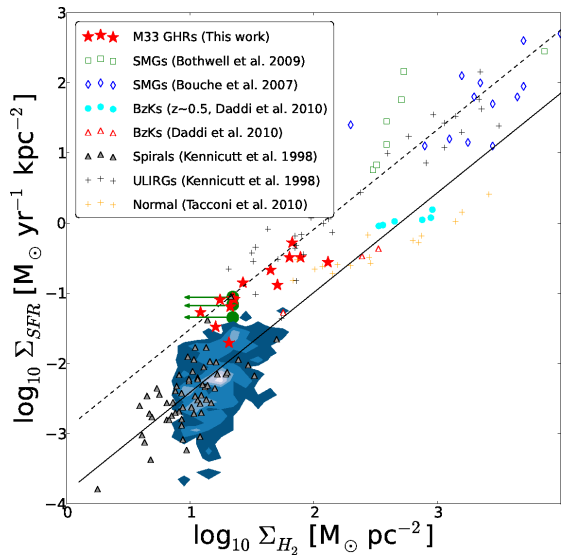}
\caption{The SFR vs molecular gas surface densities for GHRs (red filled stars, NGC604 and NGC595) and for non-GHRs (blue filled contour) in M33, compared with that of other galaxies from \citet[][and references therein]{2010ApJ...714L.118D}.
Contour levels are 1, 3, 5, 7, and 9 independent data points, per 0.05 dex cell.
Green circles represent other GHRs in M33, with the molecular gas mass estimated from the CO($1-0$) data \citep{2007ApJ...661..830R}.
\label{fig3}}
\end{figure}

Figure~\ref{fig3} shows the SF law plot of M33 together with data of other nearby and distant galaxies in D10.
The molecular gas density for M33 was derived using the following equation:
$\Sigma_{\rm H_2}[M_{\sun}\,{\rm pc}^{-2}]= 4.81 \,I_{\rm CO(3-2)}X_{\rm CO}(R\ratio)^{-1}$,
where we assume that the $X_{\rm CO}$ factor in M33 is the Galactic value, $X_{\rm CO, Gal}=(3\pm1)\times10^{20}\,{\rm cm^{-2}}({\rm K\,km\,s^{-1}})^{-1}$ \citep{1988A&A...207....1S}, and a fixed conversion factor from $I\cott$ to $I\cooz$, $R\ratio=0.4$, corresponding to the average line ratio over the M33 GMCs (Paper\,{\sc i}).
We account for Helium (a factor of $\sim1.36$) but not for the inclination of the galaxy.

We plot data for the GHR and non-GHR domains separately: data points for GHRs are shown as red filled stars, while for non-GHR as blue contours.
The solid line corresponds to the fit to normal galaxies (slope of 1.42) and the dashed line is the same relation offset by 0.9 dex to fit starburst galaxies. These represent the ``sequence of disks'' and ``sequence of starbursts'', respectively.
The method to estimate $\Sigma_{\rm H_2}$ is different in the definitions among the different galaxy samples, but are not large enough to explain the offset of 0.9 dex between the two sequences (D10). 

We find that the majority of data points in M33 are aligned along the sequence of disks, but the SFRs in GHRs are much higher and reach the sequence of starbursts. 
For reference, we also plot other bright H{\sc ii} regions in M33: NGC\,592, NGC\,588 and IC131. Their SFR is measured as in \S~2.2, but we calculate an upper limit of $\Sigma_{\rm H_2}$ from the CO($1-0$) observations in \citet{2007ApJ...661..830R} because these are outside of our CO($3-2$) mapping area.
These upper limits also lie along the starburst sequence.

In summary, the SFEs in GHRs are $\sim1$ dex higher than that in {\it normal} disk regions. 
\citet[][]{1995ApJ...455..125W} previously reported that NGC\,604 and NGC\,595 had a factor of 3 higher SFE than the average over the disk.
 The differences between our result and them might come from the different definition and method to calculate stellar masses and SFEs.
 Their definition of SFE is given as mass of optically visible stars formed per GMC mass, while ours include the contribution from embedded stars, derived from a combination of the H$\alpha$ and 24$\micron$ data.
Also, we employ pixel-to-pixel analysis, which focus on the molecular gas at the very vicinity of the star forming region.

\section{Properties of Giant Molecular Clouds Around Giant H{\sc ii} Regions}
\label{ghr}
The offset between the two sequences has been argued to be related to a different physical origin due to (1) a different $X_{\rm CO}$ factor, (2) the fraction of dense molecular gas, or (3) the effect of a top-heavy IMF in starbursts (\S\,1).
In the following subsections, we confront the peculiarity of high SFE in GHRs against the three different explanations above.

\subsection{Is There any Difference in the $X_{\rm CO}$ Factor?} 
\label{sec:xco}
A general method to derive the $X_{\rm CO}$ factor is to compare the virial masses ($M_{\rm vir}$) and the CO($1-0$) luminosity of a cloud.
In order to examine the relation between $M_{\rm vir}$ and CO($1-0$) luminosities in M33, we use the GMC catalog in \citet{2007ApJ...661..830R} because their data has a resolution high enough to resolve the typical size of a GMC ($\sim50$\,pc).
As in their catalog, we calculated the $M_{\rm vir}$ using the equation in \citet{1990ApJ...363..435W}.
The GMC sizes were derived using $D_{\rm pc} = \sqrt{ A_{\rm maj} \times A_{\rm min}}$, where $A_{\rm maj}$ and $A_{\rm min}$ are the major and minor axes. $D_{\rm pc}$ spans from 50\,pc to 160\,pc.

\begin{figure}
\plotone{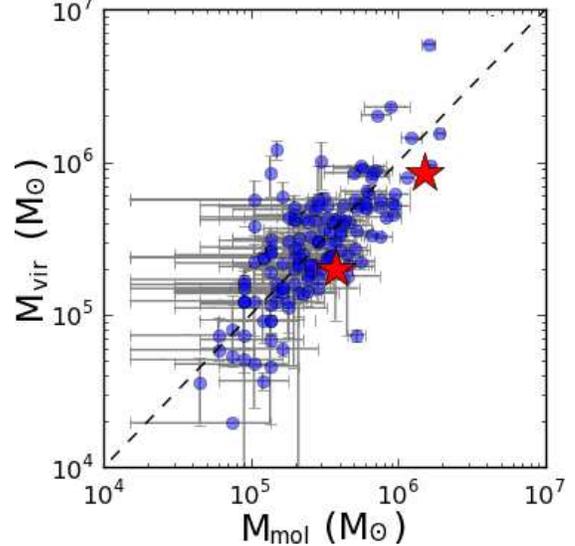}
\caption{Plot of the virial mass of the M33 GMCs as a function of the molecular mass. 
These are based on the CO(1--0) GMC catalog with a resolution of $\sim50$\,pc  \citep{2007ApJ...661..830R}.
Star symbols represent GHRs. The dashed line indicates the case where the virial equilibrium holds. 
\label{fig4}}
\end{figure}

A plot of the $M_{\rm vir}$ as a function of the molecular mass ($M_{\rm mol}$) is shown in Figure~\ref{fig4}, which is useful to derive the $X_{\rm CO}$ factor.
Here we use the $M_{\rm mol}$ from \citet{2007ApJ...661..830R}, but using $X_{\rm CO, Gal}$.
Star symbols in Figure~\ref{fig4} represent data for GHRs, NGC~595 and NGC~604.
The best fit to all data is $M_{\rm vir}=(1.03 \pm 0.08) M_{\rm mol}$ if we take absolute errors into account.
The $X_{\rm CO}$ factor for each cloud in M33 is expressed as $X_{\rm CO}=X_{\rm CO, Gal}\frac{M_{\rm vir}}{M_{\rm mol}}$ \citep{1995ApJ...448L..97W}.
Thus the fit suggests that the $X_{\rm CO}$ factor of $(3.1\pm0.2)\times10^{20}\,{\rm cm^{-2}\,(K\,km\,s^{-1})^{-1}}$ is likely the best value to estimate the $M_{\rm mol}$ assuming that the virial equilibrium holds.

If we interpret that the difference values between GMCs associated with the GHRs and ones with non-GHRs, the $X_{\rm CO}$ factor is slightly lower in these GHRs than other fields:
their virial parameter is $M_{\rm vir}/M_{\rm mol}\sim 0.55\pm0.20$, which corresponds to a $X_{\rm CO}$ factor of $(1.5\pm0.2)\times10^{20}\,{\rm cm^{-2}\,(K\,km\,s^{-1})^{-1}}$.
When this $X_{\rm CO}$ factor is applied, the calculated molecular mass for the GHRs decreases compared to the obtained value with the $X_{\rm CO, Gal}$ factor, and thus this results in an even higher SFE. 
Another interpretation is that SF clouds are more gravitationally bound that as data points for GHRs are below the virial equilibrium line.
At any cases, the $X_{\rm CO}$ factor does not explain why the SFEs in GHRs are high.

\subsection{Is SFE Higher in Dense Gas?} 
\label{sec:dense}

\begin{figure}
\epsscale{1.1}
\plotone{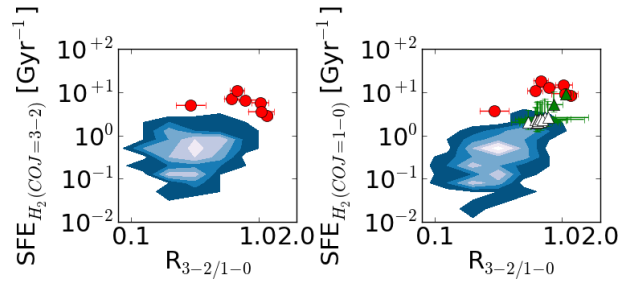}
\caption{Plot of the $R\ratio$ versus SFE$_{{\rm H_2 (CO}\,J=3-2)}$ (left) and SFE$_{\rm H_2(CO\,J=1-0)}$ (right).  Data for GHRs are shown as red points, while that for non-GHRs as blue contours. 
Contour levels are 1, 3, 5, 7 and 9 independent data points per 0.05 dex cell.
Data points where error are better than 40\% are shown.
The data points in several annuli of M83 are shown as green filled (unfilled) triangles, representing the center (disk)\citep{2007PASJ...59...43M}.
 \label{fig5}}
\end{figure}

The $R\ratio$ provides a rough estimation of the gas density ($n_{\rm H_2}$) and kinetic temperature ($T_{\rm kin}$) \citep[e.g.,][]{2008ApJS..175..485M}.
Figure~\ref{fig5} shows the plot of the $R\ratio$ versus SFE derived from CO($3-2$) and CO($1-0$) (hereafter SFE$_{{\rm H_2(CO}\,J=3-2)}$ and SFE$_{{\rm H_2(CO}\,J=1-0)}$).
The CO($1-0$) molecular gas surface densities are calculated using $\Sigma_{\rm H_2}= 4.81I_{\rm CO(1-0)}X_{\rm CO,Gal}$.

In the SFE$_{{\rm H_2(CO}\,J=3-2)}$ - $R\ratio$ plot, we find that the non-GHRs data points show a nearly flat distribution,
which is consistent with other studies \citep{2010ApJ...714..571W}.
On the contrary, a positive correlation is apparent in the SFE$_{{\rm H_2(CO}\,J=1-0)}$ - $R\ratio$ plot.
This trend is similar to that in the inner kiloparsec of M83, known to host a starburst nucleus \citep{2007PASJ...59...43M}.
When taking it into account that $R\ratio$ can be a diagnostic of dense gas fraction at GMC scale \citep{2008ApJS..175..485M}, these plots suggest that the SFE of the dense molecular gas is independent of the dense gas fraction, while the SFE of the more diffuse molecular gas is enhanced in dense gas regions.

In both plots the SFE and $R\ratio$ of GHRs are higher than in non-GHRs.
Their $R\ratio\sim1$ suggests that these GMCs are dense and warm.
We calculate how different the physical properties between GHRs and non-GHRs are according to the large velocity gradient (LVG) approximation, by using the $^{12}{\rm CO}(1-0)/^{13}{\rm CO}(1-0)$ ($R_{13/12(J=1-0)}$), $^{12}{\rm CO}(2-1)/^{13}{\rm CO}(2-1)$ ($R_{13/12(J=2-1)}$) and $^{12}{\rm CO}(3-2)/^{12}{\rm CO}(2-1)$ ratios ($R_{3-2/2-1}$) of seven GMCs in \citet{1997ApJ...483..210W}.
This follows the same prescription as in  \citet{2008ApJS..175..485M}, which was used for the LMC where the metallicity is similar to that of M33.
\citet{1997ApJ...483..210W}'s observations have a similar beam size except $^{13}$CO(1--0) observations (55\arcsec),
and thus their $R_{13/12 (1-0)}$ is calculated with the resolution of $^{13}$CO(1--0) data.
All seven GMCs are covered by our CO(3--2) observations and named in Paper\,{\sc i}
as GMC-1, 5, 15, 17, 18, 27, and 72.
Note that their alternative names in \citet{1997ApJ...483..210W} are NGC604-2, MC1, MC32, MC19, MC13, NGC604-4 and MC20, respectively.

\begin{figure}
\epsscale{1.1}
\plotone{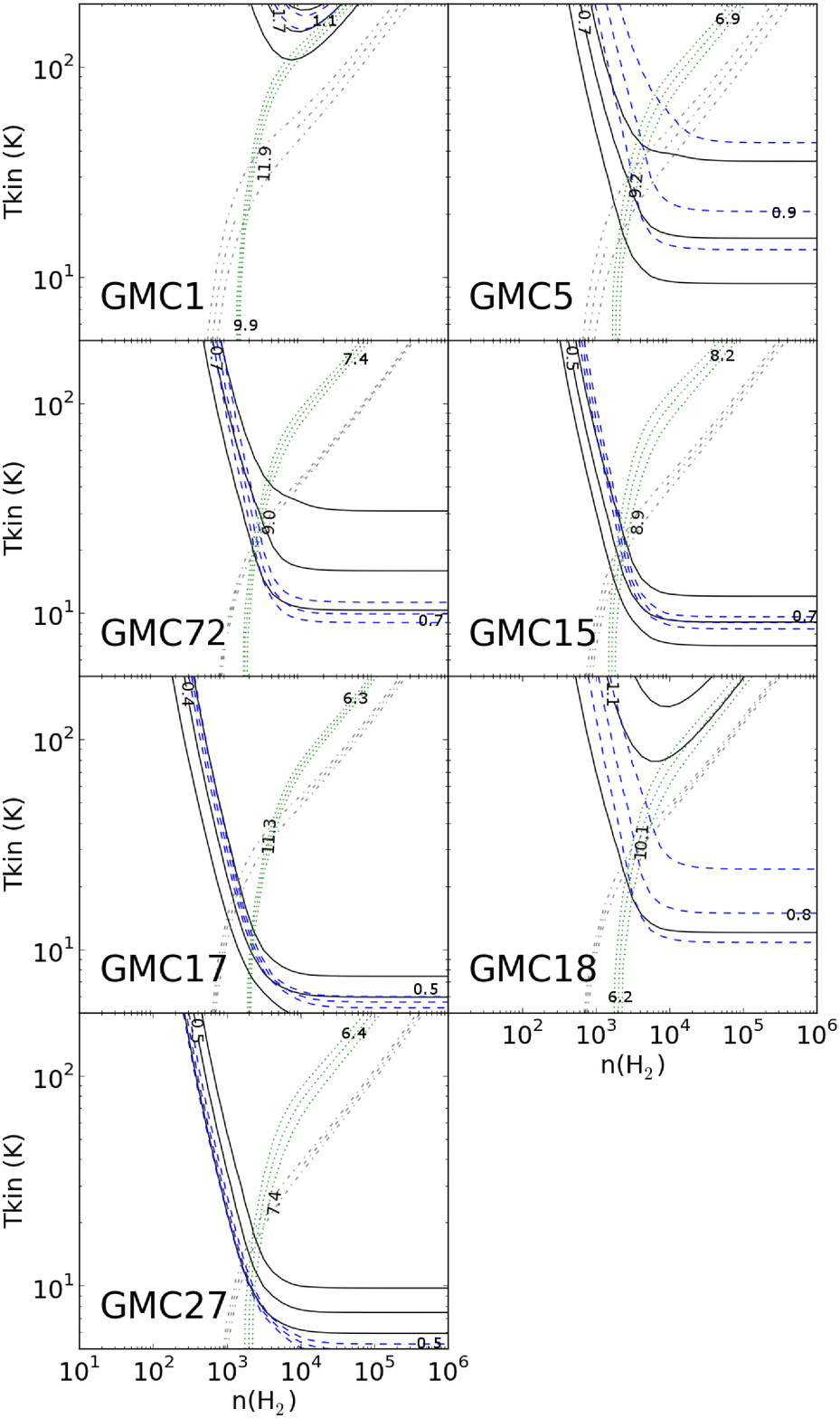}
\caption{Kinetic temperature ($T_{\rm kin}$) vs molecular hydrogen density ($n({\rm H}_2)$) plots for seven GMCs, using the LVG approximation. 
Gray dotted-dashed, green dotted, blue dashed and black solid lines indicate indicate $R_{13/12(J=1-0)}$, $R_{13/12(J=2-1)}$, $R_{3-2/2-1}$ and $R\ratio$ ratios, respectively.
The values of each ratio are labeled in their corresponding lines.
The three lines for each ratio indicate the line ratio with the intensity calibration errors ($\pm\,1\sigma$).
Acceptable values for $T_{\rm kin}$ and $n({\rm H}_2)$ are the region in which these four lines overlap within the intensity calibration errors.\label{fig6a}}
\end{figure}

The $R\ratio$ of the corresponding position for the seven GMCs are measured from the CO($3-2$) and CO($1-0$) maps of the individual GMC in Paper\,{\sc i}.
In Figure\,\ref{fig6a} we present the LVG analysis from the four line ratios $R\ratio$ (black solid line), $R_{13/12(J=1-0)}$ (gray dotted-dashed line), $R_{13/12(J=2-1)}$ (green dotted line) and $R_{3-2/2-1}$ (blue dashed line).
Since the lines for $R\ratio$ and $R_{3-2/2-1}$ ratios well overlapped on the LVG plane, 
in general our $R\ratio$ measurements are consistent with $R_{3-2/2-1}$ in \citet{1997ApJ...483..210W}.
However, we cannot obtain solutions for GMC-17 and GMC-1 (NGC\,604) with these four line ratios \citep[even with the data of ][]{1997ApJ...483..210W}.
Because the $R_{13/12(J=1-0)}$ ratio traces the properties of a larger portion of the molecular cloud, this cause large uncertainty due to averaging and we preferred to use $R_{13/12(J=2-1)}$ in the analysis rather than $R_{13/12(J=1-0)}$.
If we exclude the $R_{13/12(J=1-0)}$ ratios in GMC-17 and GMC-1 we obtain $n_{\rm H_2}\sim(1-2)\times10^3\,$cm$^{-3}$ and $T_{\rm kin}=10-20$\,K for non-GHRs (5, 15, 17, 18, 27, and 28), while $n_{\rm H_2}>2\times10^4\,$cm$^{-3}$ and $T_{\rm kin}>100$\,K for GHR (GMC-1).

In summary, the GMCs in GHRs are about 10--20 times denser and warmer than the ones in non-GHRs.
Since free-fall time is inversely proportional to the density square root, the dense molecular gas would tend to collapse
to form stars rapidly, which would result in a higher SFE : SFE\,$\propto$ SFR/($M_{\rm gas}$/$\tau_{\rm ff})$ \citep{2012ApJ...745...69K}. 
For densities three times larger, the SFE would increase by a factor of 3--4.
Therefore the high SFE in GHRs can be partly due to a larger local dense gas fraction.

\subsection{Top-heavy IMF in GHRs}
The IMFs in NGC~604 and NGC~595 are previously found to be flatter \citep[$\alpha=1.88$ and 1.92 \footnote{The IMF is expressed as $dN\propto m^{-\alpha}dm$, where $dN$ is the number of stars with masses in a range of $m$ to $m+dm$. };][]{1993AJ....105.1400D} than the Salper IMF ($\alpha=2.35$), similar to those observed in starburst galaxies \citep[e.g.,][]{1980ApJ...238...24R}.
We estimated the SFR with different IMFs for the GHRs and the disk, using the stellar population models of Starburst99\footnote{See \url{http://www.stsci.edu/science/starburst99}} \citep{1999ApJS..123....3L}.
For non-GHRs we have used the Starburst99 default parameters of a constant SF model, which consists of a Salpeter IMF with mass limits of 0.1--100\,$M_{\sun}$, solar metallicity $Z_{\sun} = 0.02$ and a 100\,Myr duration model of constant SF.
For GHRs we adopted the same parameters except for the IMF.
The conversion from the (extinction-free) H$\alpha$ luminosity, $L({\rm H}\alpha)$ [${\rm erg\,s^{-1}}$], to SFR [$M_{\sun}\,{\rm yr}^{-1}$], with a Salpeter IMF is ${\rm SFR}= 6.31 \times 10^{-42} L({\rm H}\alpha)$ in case of IMF of 1.88, and 
${\rm SFR}= 1.76 \times 10^{-42} L({\rm H}\alpha) $,
and in case of IMF of 1.92, ${\rm SFR} = 1.91 \times 10^{-42} L({\rm H}\alpha) $.
The SFRs derived with a flatter IMF are a factor of 3--4 smaller than those with a Salpeter IMF.
This result does not considerably change even if we use the starburst model in the Starburst99 instead of constant SF model.
Therefore, a flatter IMF in GHRs may contribute to the SFE's 1 dex offset.

\section{Implications}
\label{sec6}
The two different regimes on the SF law plot of GHRs and non-GHRs may arise from the contribution of a different fraction of dense molecular gas and a top-heavy IMF, but not from a different $X_{\rm CO}$ factor.
In this section, we provide a scenario in which the combination of these two factors may explain the high SFE and $R\ratio$ found in GHRs.

A top-heavy IMF is characterized by an overabundance of massive stars.
Once massive stars are born in a cluster, they produce intense ionization photons and stellar winds, and then finally become supernovae. 
These radiative and mechanical energetic input can destroy
the structure of the parental molecular clouds, which may suppress subsequent SF (negative feedback). 
Contrary to this, the shock front emerges and compresses the surrounding gas into a very dense layer where second SF is triggered (positive feedback).
SF regions where positive feedback overcome the negative one has been witnessed in M33 as well as in our Galaxy \citep[e.g.,][]{2012A&A...538A..11B}.
In the two GHRs, NGC\,604 and NGC\,595, dense molecular gas was detected at the periphery of the H{\sc ii} regions where massive stars are born but molecular gas in its center is scarce \citep[e.g.,][]{2007ApJ...664L..27T,2010ApJ...724.1120M,2009ApJ...699.1125R}.

According to \citet{2006ApJ...646..240H}, the gas swept up to the layer around an expanding H{\sc ii} region ($M_{\rm sh}$) is proportional to the ionizing photon rate ($Q^{\ast}$) when a constant initial ambient gas density is assumed. 
We calculated the $Q^{\ast}$ for the 18 M33 YSGs, whose ages are estimated to be less than 10\,Myr by counting the amount of O stars from our YSG catalog (Paper\,{\sc i}), and assuming a standard value of $Q^{\ast}$ for each stellar type \citep{2005A&A...436.1049M}.
The calculated $Q^{\ast}$ per YSG spans (1--$39)\times10^{50}$\,photons\,s$^{-1}$, and among which those in the two GHRs are at least 2--30 times larger than in other H{\sc ii} regions. 
In derivation of the $M_{\rm sh}$ from the $Q^{\ast}$, the scaling relations introduced in equations (39)--(41) of \citet{2006ApJ...646..240H} is used
and a classical H{\sc ii} region created by a single star with a mass of $100\,M_{\sun}$ is adopted as a standard model \citep{2006ApJ...646..240H}.
In the uniform ambient density of $10^3\,{\rm cm}^{-3}$, the molecular gas of (3 -- 18)$\,\times10^5\,M_{\sun}$ can be accumulated in the shell within 3--4 Myr in the two GHRs. 
This is 2--30 times larger in the GHRs than in other H{\sc ii} regions.

This estimation is for the case of star formation under an uniform ambient density.
The estimated $M_{\rm sh}$ depends in practice on physical conditions such as the ambient (pre-existing) density structure, stellar types and distribution \citep{2006ApJ...646..240H}.
For instance, the GHRs are characterized by complex distributions such as filaments, shells and bubbles \citep[``champagne flow'';][]{2006ApJ...643..186T}.
In these
cases, the ionization front rapidly erodes the parental cloud and only a part
of the mass is swept-up and remains within the shell, which might result
in smaller $M_{\rm sh}$ \citep{2005ApJ...623..917H}.
Although the use of refined numerical simulations is needed, this suggests that a larger amount of molecular clouds is accumulated in GHRs than in normal H{\sc ii} regions, which shields the FUV radiation field.
In this way the subsequent SF occurs efficiently in a short time scale and would result in such high SFE in GHRs.

Although we have focused only on feedback via the expanding H{\sc ii} region, we note that the material for new stars can survive and the gas densities become larger even if the stellar feedback via supernova explosion is considered.
For example, 30\,Doradus is the brightest GHR in LGGs and many SNe reside.
On the other hand, no candidates of SN remnant has been found at the vicinity of the central clusters of NGC\,604 \citep{2005AJ....130..539G}.
The 30\,Doradus nebula is associated with CO molecular clouds with a total mass of $\sim4\times10^5\,M_{\odot}$ at the ridge of the central cluster \citep{1998A&A...331..857J}, in which the fragmentation has occurred to form dense clumps and consequently new stars \citep{2013ApJ...774...73I}.
These molecular clouds have survived against the strong radiation from the central cluster and might be compressed by the pressure of the warm H{\sc ii} gas or the hot gas generated by shock heating from stellar winds and SNe \citep{2011ApJ...738...34P,2011ApJ...731...91L}.
The swept-up molecular clouds could survive possibly because each fragment contracts to a dense clump, the column density increases and molecules will be protected against strong radiation \citep{2007ApJ...664..363H}.

\section{Summary}
\label{indiv:sum}
In this Paper, we have studied the SF law and SFEs in the GHRs and other disk regions of M33, based on wide field and high sensitivity CO($3-2$) and CO($1-0$) observations at 100\,pc resolution. 
We have examined three possibilities to contribute to the high SFEs in GHRs, such as $X_{\rm CO}$ factor, IMF and dense molecular gas fraction. 
Our results are the following:
\begin{enumerate}
\item We found high SFEs and $R\ratio$ in the molecular clouds around GHRs. SFEs differs about one dex from that of other disk regions in M33. Such high SFEs in GHRs is comparable to that in more distant starburst systems.
\item We examined a possible variation of the $X_{\rm CO}$ factor between the GHRs and other disk regions to explain that the $X_{\rm CO}$ factor does not contribute to the high SFE in M33 GHRs. We used the relation between virial masses and molecular masses and found that the estimated $X_{\rm CO}$ factor of GHRs are lower than that of other regions by a factor of 2. 
\item The correlations between $R\ratio$ and SFE show a clear positive correlation. This suggests that the dense gas fraction traced by $R_{\ratio}$ is correlated to SFE. 
To quantify this, we also conducted the LVG analysis for seven M33 GMCs including a GMC associated with GHR and showed that the densities in GHRs are a factor of 3--4 larger than the other GMCs associated with normal H{\sc ii} regions.
This suggests that the variation in dense gas fraction can partially explain the high SFE in GHRs.
\item The SFR is calculated by adopting a top-heavy IMF in GHRs while a Salpeter IMF in other regions, using the Starburst99 program, which resulted in a difference of a factor of 3--4. This suggests that a flatter IMF in GHR than in other regions partially contributes to the SFE's 1 dex offset.
\item We conclude that the high SFE at the GHRs can be interpreted by a combination of a different IMF and a larger fraction of dense molecular gas, but not due to a different $X_{\rm CO}$ conversion factor.
\item We suggest a scenario that the parental molecular gas would tend to get denser in the accumulated gas around the first generation stars, while to be rapidly consumed by SF, eroded and dissipated due to more massive stars in GHRs. This results in high SFE and densities.
We also suggest that this scenario localized around GHRs can be applied to the distant starburst systems where lack of resolution prevent us from resolving the building elements of the starburst. 
\end{enumerate}

\acknowledgments
We gratefully acknowledge the contributions of the ASTE staff to the development and operation of the telescope.
The ASTE project is driven by NRO, a branch of NAOJ, in collaboration with University of Chile, and Japanese institutes including University of Tokyo, Nagoya University, Osaka Prefecture University, Ibaraki University, Hokkaido University and Joetsu University of Education.


\end{document}